 \documentclass[10pt,conference]{IEEEtran}
\IEEEoverridecommandlockouts
\usepackage{cite}
\usepackage{authblk}
\usepackage{amsmath,amssymb,amsfonts}
\usepackage{algorithmic}
\usepackage{graphicx}
\usepackage{textcomp}
\usepackage{xcolor}
\usepackage{multirow}
\usepackage{balance}
\usepackage{lscape}
\usepackage{multicol}
\usepackage{rotating}
\usepackage{amsmath}
\usepackage{bbding}
\usepackage{amssymb}
\usepackage{makecell}
\usepackage{pifont}
\RequirePackage{pdflscape}
\def\BibTeX{{\rm B\kern-.05em{\sc i\kern-.025em b}\kern-.08em
    T\kern-.1667em\lower.7ex\hbox{E}\kern-.125emX}}

\usepackage{tcolorbox}

\newcommand{\find}[1]{
\begin{tcolorbox}[leftrule=0.5mm,toprule=1mm,bottomrule=1mm,left=0pt,right=0pt,top=2pt,bottom=2pt] 
#1
\end{tcolorbox}
} 

\usepackage{hyperref}
\hypersetup{
     colorlinks = true,
     linkcolor = blue,
     anchorcolor = red,
     citecolor = magenta,
     filecolor = blue,
     urlcolor = black,
}

\begin{document}

\title{Privacy in Foundation Models: A Conceptual Framework for System Design}

\author{Tingting~Bi$^1$, 
        Guangsheng~Yu$^2$,
        Qin~Wang$^3$
        }
\affil{
$^1$\textit{The University of Melbourne, Australia}\\ 
$^2$\textit{University of Technology Sydney, Australia}\\ 
$^3$\textit{CSIRO Data61, Australia}
}

\renewcommand*{\Affilfont}{\normalsize}


\maketitle
\begin{abstract}
AI and its relevant technologies are currently undergoing a profound transformation of development and organizational processes within companies. Foundation models present both significant challenges and incredible opportunities. In this context, ensuring the quality attributes is of importance, and with a particular focus on the challenging issue of privacy due to the sensitive nature of the data and information involved.

However, there is currently a lack of consensus regarding the scope of both technical and non-technical issues that the privacy evaluation process should encompass. Additionally, there is uncertainty about which existing methods are best suited to effectively address these privacy concerns. In response to this challenge, this paper introduces a novel conceptual framework that integrates various responsible AI patterns from multiple perspectives, with the specific aim of safeguarding privacy.

\end{abstract}

\section{Introduction}\label{sec_intro}

The use of AI by corporations and governments has grown exponentially, and this growth has been predetermined by many abilities of AI, such as analysis of large amounts of data, predictions with high accuracy rates, identification of trends and patterns, predictions of complex associations, improvement of profitability, analysis of financial rations/risks.

\textit{\textbf{Foundation models}} (FMs) have become indispensable (i.e., foundational components) for various AI systems due to their ability to learn from vast amounts of data and adapt to a wide range of tasks \cite{bommasani2021opportunities}. Foundation models are pre-trained on extensive data sets and can be fine-tuned to perform specific tasks, making them highly versatile and efficient \cite{mai2023opportunities}. 
Incorporating foundation models as components of software systems holds great opportunities \cite{mai2023opportunities}.

During AI system development, many different stakeholders make many complex design decisions, and the use of AI and foundation models raises a set of Quality Attribute (QA) concerns. These QAs serve as criteria to assess the quality of systems and alignment with business goals, thus influencing the selection of design approaches and shaping the structure and behavior of software systems.
\textbf{\textit{Privacy}}, being one of the paramount QAs demands heightened attention from stakeholders due to its significant implications for both individuals and society in terms of potential harm\cite{anthonysamy2017privacy}. There is a set of privacy concerns in both classical and foundation model-based systems, for example, ensuring the security of data used to train and fine-tune foundation models is crucial, and data breaches or unauthorized access to the data can lead to significant privacy violations. 

As such, it is crucial to implement data sanitation and strict user policies to mitigate such privacy issues. Some potential solutions, for example, involve for consumers of foundation model-based systems to be aware of how to safely interact with foundation-model-based systems and identify the risks associated with unintentionally inputting sensitive data that may return by the systems in output elsewhere \cite{cheng2023prompt}. 

Many factors should be considered and many decisions can be made when building an AI system. Software architecture of AI systems undergoes constant change as users incessantly demand more functionality, and adapting to a dynamic environment is often crucial for a system's success \cite{lindvall2003empirically}. Software architecture evolves over time in response to these changes, and achieving successful software evaluation poses a significant challenge. The key to seamless software evaluation lies in the system's structure. Furthermore, several problems challenge the privacy design of AI systems due to information and  knowledge availability \cite{lecue2020role}. \textbf{\textit{Responsible AI (RAI) patterns}} are reusable solutions to a problem commonly occurring within a given context in software development. A set of solutions contribute to multiple responsible AI  principles \cite{lu2023towards}. In this work, we focus on the RAI patterns that practitioners and stakeholders can undertake to ensure privacy in foundation-model based systems at different levels.

\textbf{Scope of the proposed framework}: We propose a conceptual framework and guideline, which incorporate (select) RAI patterns for making decisions to address privacy concerns. To be specific, the conceptual framework provides a structured and systematic process, from different perspectives, to address and protect user data, and comply with legal and ethical privacy requirements. We endeavor to create an enhanced iteration that incorporates responsible design principles essential for addressing a set of QA requirements. 

\textbf{Our contributions}. We make the following contributions: 

\begin{itemize}
    \item We introduce a systematic process incorporating various \textit{RAI patterns}, specifically focusing on FM-based software systems, placing a strong emphasis on \textit{privacy}. This process equips stakeholders with the tools to better understand how privacy can be ensured within FM-based systems from various perspectives.

    \item  Our conceptual framework has broad applicability. The framework allows stakeholders to assess risks and make informed decisions. It plays a crucial role in identifying recurring patterns or strategies within the system planning and design process. Additionally, it assists in forecasting maintenance costs associated with addressing specific concerns, such as estimating anticipated maintenance expenses over a defined time period.

    \item We introduce an \textit{RAI patterns pool}, which can be generically applied for evaluating other QAs in foundation model-based systems. This pool takes into consideration stakeholder expectations, ethical parameters, and risk assessment. The evaluation process is to make sure transparency and trustworthy and is enriched with practical insights derived from real-world industry scenarios.

\end{itemize}

\vspace{0.5cm} 


\section{Background and Related Work}\label{sec_relatedwork}

In this section, we discuss the related work from two aspects: (1) previous work on foundation model-based software systems; 
and (2) privacy design in software engineering.

\subsection{Foundation Model-based Software Systems}
\label{subsec_relatedWork_foundationModel}

\begin{figure*}[t]
\centering
\includegraphics[width=0.9\textwidth]{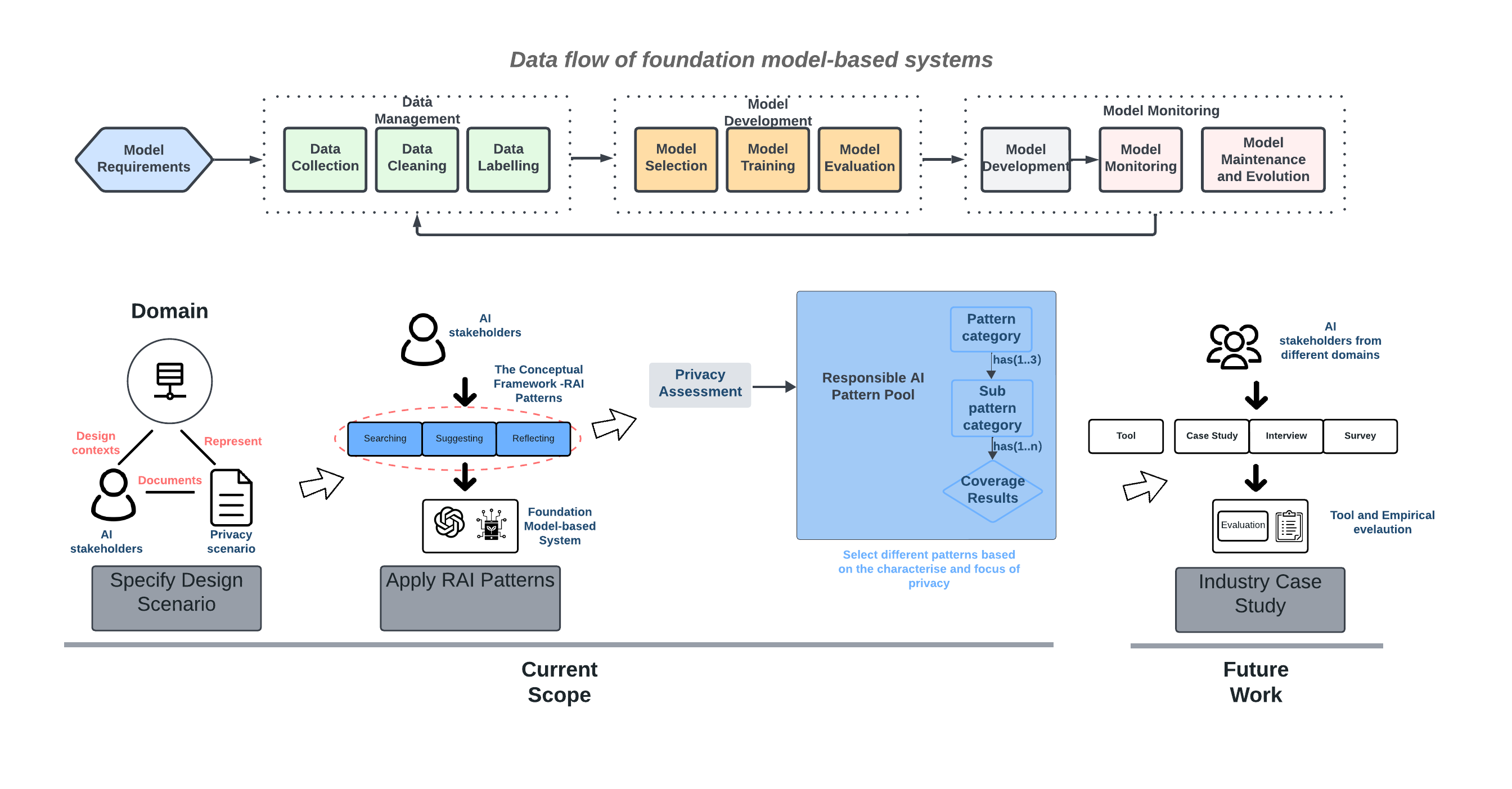}
\vspace{-1cm}

\caption{Data flow and incorporating RAI patterns in foundation-based systems}
\centering
\label{FoundationModel_based_Systems}
\end{figure*}

Existing studies explored the utility of foundation models in software systems; for example, Xing \textit{et al.} discussed AI chain engineering through Prompt Sapper~\cite{cheng2023prompt}.  Terry \textit{et al.} proposed an evaluation framework for code generation using GPT-3.5 Turbo, improving upon CodeBERTScore~\cite{code-generator}. LangChain provided a framework for data-aware applications driven by language models~\cite{LangChain2023}. AutoGPT illustrated the versatility of foundation models in tasks like code editing and internet-based tasks~\cite{Auto-GPT2023}. Qin \textit{et al.} delved into tool learning and its synergy with foundation models~\cite{qin2023tool}. Novel contributions have also been made in non-English literature. Methods using language outputs from Large Language Models (LLMs) for robotic skill synthesis have been proposed~\cite{yu2023language}. Strategies enabling LLMs as autonomous decision-makers in autonomous driving have been introduced~\cite{fu2023drive}. Shepherd is a critic for evaluating generated text by other large models~\cite{wang2023shepherd}. LLMs have been used for efficient database administration~\cite{zhou2023llm}. Lastly, a closed-loop framework has been put forward where LLMs create reusable tools for problem-solving~\cite{cai2023large}.

\subsection{Privacy Goals and Patterns}

\label{relatedWork_privacy}

This work leverages the conceptual framework to ensure the different privacy goals. Table~\ref{tab:privacy_goals_and_issues} illustrates the relationship between privacy goals and tactics  \cite{anton2002analyzing}. Privacy principles and goals come from different legislation (e.g., GDPR, LGPD, PIPEDA) and privacy standards (e.g., ISO31700, ISO29100, and NIST privacy framework). 

To make the appropriate decisions, it is important to determine whether the desired protection goal relates to \emph{unlinkability}, \emph{intervenability}, or \emph{transparency}.  
We first discuss tactics and  privacy patterns (see Table \ref{tab:privacy_goals_and_issues}): \textit{1) Increase cohesion (IC)}: this tactic reduces central data collection;
\textit{2) Reduce the size of a module (RS)}, this tactic is about the modification costs that could be reduced when changing a smaller module as any future changes;  \textit{3) Use of intermediary (UI)}: the use of intermediary tactics in privacy design involves implementing intermediaries or middle layers in systems, applications, or processes to enhance privacy protection and control over personal data; \textit{4)} \textit{Encapsulate (EN)}: this tactic involves isolating or enclosing sensitive data and privacy-related functionality within a well-defined and controlled environment; \textit{5) Abstract common services (AC)}: enables users to abstract their data access control on several service providers and move to the common module on the user side; \textit{6)} \textit{Restrict dependency (RD)}: this tactic involves minimizing the reliance on external libraries, third-party services; \textit{7): Defer binding (DB):} involves delaying the binding or association of personal data with specific individuals or entities for as long as possible within a system or process. In the following section, we introduced the privacy goals that are related to privacy design tactics.

\textbf{Unlinkability} can be achieved by data oriented patterns and tactics. It ensures that personal data is not misused and users know how their data will be utilized. The primary focus here is to inform users about the privacy practices of organizations~\cite{10.1145/3411764.3445516}.
The subgoals can be undermined by information hide, where invasive surveillance infringes upon a user's right to be informed about their data usage. 
There is a set of guidelines to make it concrete: 1) collecting or copying unnecessary attributes to the dataset if this is irrelevant for the purpose; 2) anonymizing the data where possible; 
3) delete the data as soon as possible when it is no longer useful (e.g., data from 7 years ago may not be relevant for the model); and 4) Minimize the number of stakeholders who accesses the data on a ``need to know" basis.


\textbf{Intervenability} can be achieved by process-oriented tactics, incorporating control into the privacy design, and requires a holistic approach that considers legal, ethical, and technical aspects ~\cite{doi:10.12968/bjon.2003.12.1.10999}. 
Different control metrics can be used, which range from group control, false positive error rate, unawareness, and counterfactual control. The relative tactics to this pattern are ``\textit{Use an intermediary}" and ``\textit{Restrict dependency }".

\textbf{Transparency} can be achieved by process-oriented tactics. It refers to the practice of making data processing activities and privacy-related practices, and decisions open and easily understandable to individuals whose data is being processed \cite{anthonysamy2017privacy}.  
This category aligns with Abstract common services.

\section{A Conceptual Framework for PRIVACY}\label{sec_methodology}


In this section, we present a conceptual framework incorporating various RAI patterns to evaluate \textbf{\textit{privacy}} in foundation model-based systems. We first present the structure of foundation model-based systems, and based on the structure \cite{lu2023responsible,lu2023operationalizing}, we subsequently present the conceptual framework.

\subsection{Architecture of Foundation Model-based Systems }

\label{subsec_foundationBase}

The concept of foundation model-based systems stems from the idea that a single powerful language model can be fine-tuned and adapted to perform a wide range of specific tasks instead of building separate specialized models for each task. By using a foundation model as a starting point, developers can take advantage of their pre-trained knowledge and language understanding to achieve better performance on specific tasks with less training data and development effort. For example, a foundation model-based system can be divided into three parts, which can be told apart and closely related to each other \cite{lu2023towards} (see Fig. \ref{FoundationModel_based_Systems}). Subsequently, it can be fine-tuned on a specific data set to create a chatbot capable of engaging in natural language conversations with users. Another use case could involve training the model on a large corpus of medical literature to develop a system that can answer medical queries or assist in diagnosing diseases. The advantage of foundation model-based systems lies in their ability to generalize knowledge and adapt to various domains and tasks. With a strong foundation model, these systems can offer better contextual understanding, generate coherent responses, and handle a wide range of inputs, and they can be applied in customer support, content generation, virtual assistants, language translation, etc~\cite{welivita2020taxonomy}.

\begin{table}[!t]
\renewcommand\arraystretch{1}
\centering
\label{table_privacygoal}
\scriptsize
\caption{Interrelation of Privacy Goals and Privacy Issue Taxonomy (IC: Increase cohesion; RS: Reduce the size of a module; UI: Use an intermediary; EN: Encapsulate; AC: Abstract common services; RD: Restrict dependency; DB: Defer binding)}
\resizebox{\linewidth}{!}{
\begin{tabular}{|c|c|m{1.8cm}|c|c|c|c|c|c|c|c}
\hline
\textbf{Goals} & \textbf{Category}& \textbf{Sub-category} & \textbf{IC}& \textbf{RS} & \textbf{UI}& \textbf{EN }& \textbf{AC} & \textbf{RD} & \textbf{DB}\\
\hline
\hline
 \multirow{4}{*}{\textbf{}} &  Minimise & 
 
 \begin{itemize}
     \item Exclude
     \item Select
     \item Strip
     \item Destroy
    \vspace{-0.3cm}
 \end{itemize}

 & \ding{52} &\ding{52}  &  & \ding{52}  & & &\\
\cline{2-10}

\textbf{Unlinkability} & Hide & 

\begin{itemize}
    \item Restrict
    \item Mix
    \item Obfuscate
    \item Dissociate
        \vspace{-0.3cm}
\end{itemize}

& & & & \ding{52}  & \ding{52}  & &\ding{52} \\
\cline{2-10}

  & Separate & 
  \begin{itemize}
      \item Distribute
      \item Isolate
          \vspace{-0.3cm}
  \end{itemize}
  
  &   & \ding{52}  & & & & \ding{52}  &\\
\cline{2-10}

  & Abstract & 
  \begin{itemize}
      \item Summarize
      \item Group
          \vspace{-0.3cm}
  \end{itemize}
  
  &   & \ding{52}  & & & & \ding{52}  &\\
\hline

{\textbf{Intervenability}} & Control &

\begin{itemize}
    \item Consent
    \item Choose
    \item Update
    \item Retract
        \vspace{-0.3cm}
\end{itemize}

&  & & \ding{52}  & & &\ding{52}  &\\
\hline

 \multirow{3}{*}{\textbf{}}  & Inform & 

 \begin{itemize}
     \item Supply
     \item Notify
     \item Explain
         \vspace{-0.3cm}
 \end{itemize}
 
 &  & & & & & \ding{52}  &\ding{52} \\
\cline{2-10}

\textbf{Transparency} & Enforce & 

\begin{itemize}
    \item Create
    \item Maintain
    \item Uphold
        \vspace{-0.3cm}
\end{itemize}

&  & \ding{52}  & & \ding{52}  & & & \\

\cline{2-10}

& Demonstrate & 

\begin{itemize}
    \item Audit
    \item Log
    \item Report
        \vspace{-0.3cm}
\end{itemize}

& & &  & & \ding{52}  & & \\
\hline
\end{tabular}
}
\label{tab:privacy_goals_and_issues}
\end{table}

\begin{figure*}[t]
\centering
\includegraphics[width=13cm]
{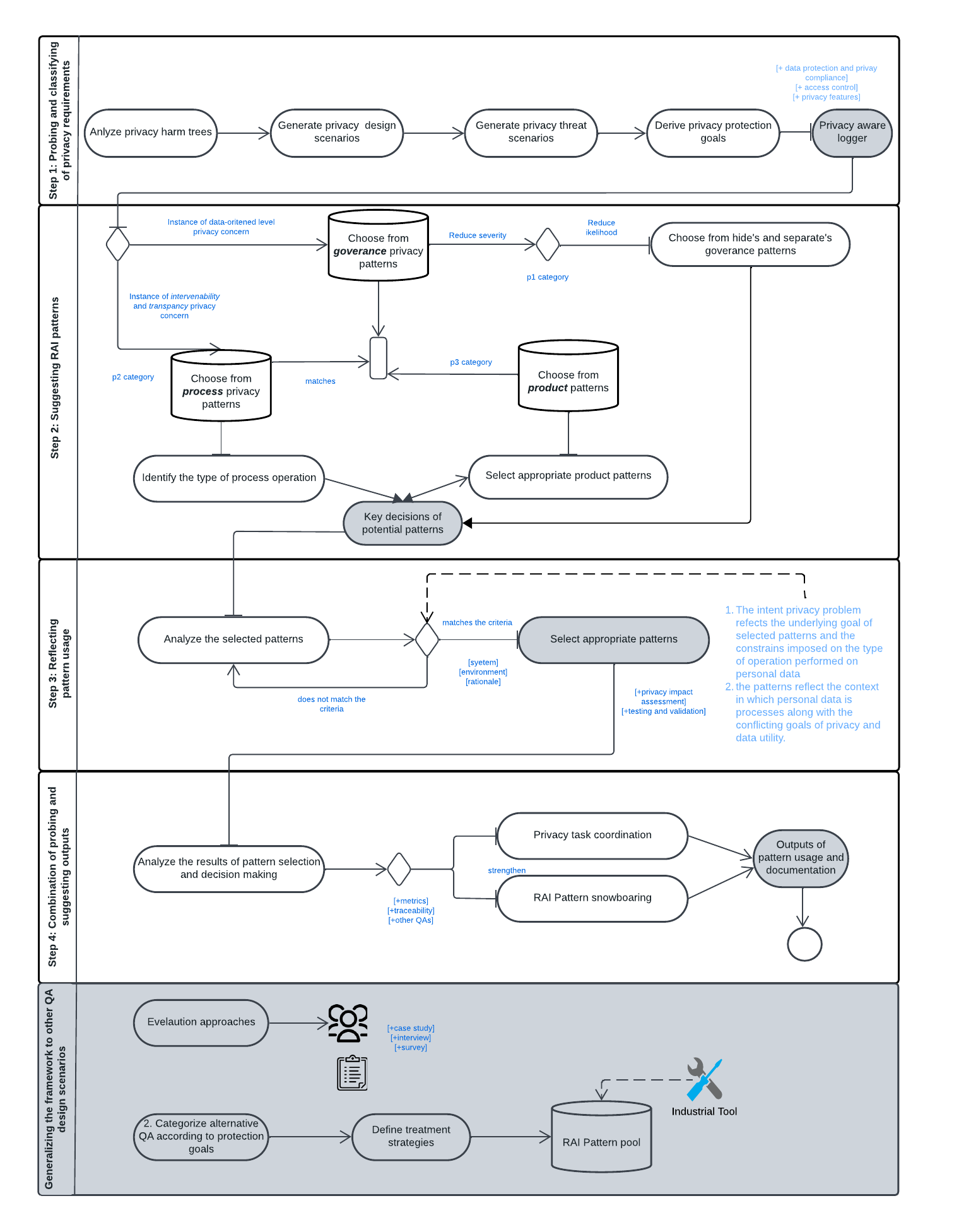}
\vspace{-0.5cm}
\caption{Conceptual framework for ensuring privacy in foundation model-based systems}
\centering
\label{Our_Framework}
\end{figure*}

From a data flow from a privacy perspective that spans the entire life cycle of foundation model-based systems (see Fig.~\ref{Our_Framework}), we can summarize the architecture as follows. 

\begin{enumerate}
    \item \textit{Data management} includes: data being collected; data being appropriately protected with best security practices; data being cleaned (e.g., pre-processing and augmentation where appropriate); and data being reported.
    \item \textit{Model development} includes: an AI model being trained on a dataset; an AI model being verified in its performance on a test dataset; and an AI model being reported.
    \item \textit{Model monitoring} includes: a model being deployed in a real-life setting with stakeholder engagement and user-centered design; updates and ongoing validation; and supervision and auditing.  
\end{enumerate}

\subsection{The Conceptual Framework}

\label{subsec_conceptual_framework}

\subsubsection {Evaluation strategy}

We present the conceptual framework as a high-level solution to capture, reason, and solve privacy issues. Our framework (Fig.~\ref{Our_Framework}) consists of three strategies to ensure privacy in FM-based systems through:

\begin{itemize}
    \item \textbf{Searching}: searching the relevant patterns in the pool and presenting questions to clarify the state of design and enquire privacy contexts.
    \item \textbf{Suggesting}: presenting potential areas of interest at conceptual level. This stage is structured to ensure comprehensive coverage of key components and features with a clear focus on privacy. 
    \item \textbf{Reflecting}: reflecting hints and questions identified patterns in the previous steps that ensure privacy in foundation model-based systems. 
\end{itemize}

The novelty of our framework is a holistic approach to ensure privacy and can be generalized for other QAs.

\subsubsection{RAI patterns}

\begin{figure*}[t]
\centering
\includegraphics[width=16.5cm]
{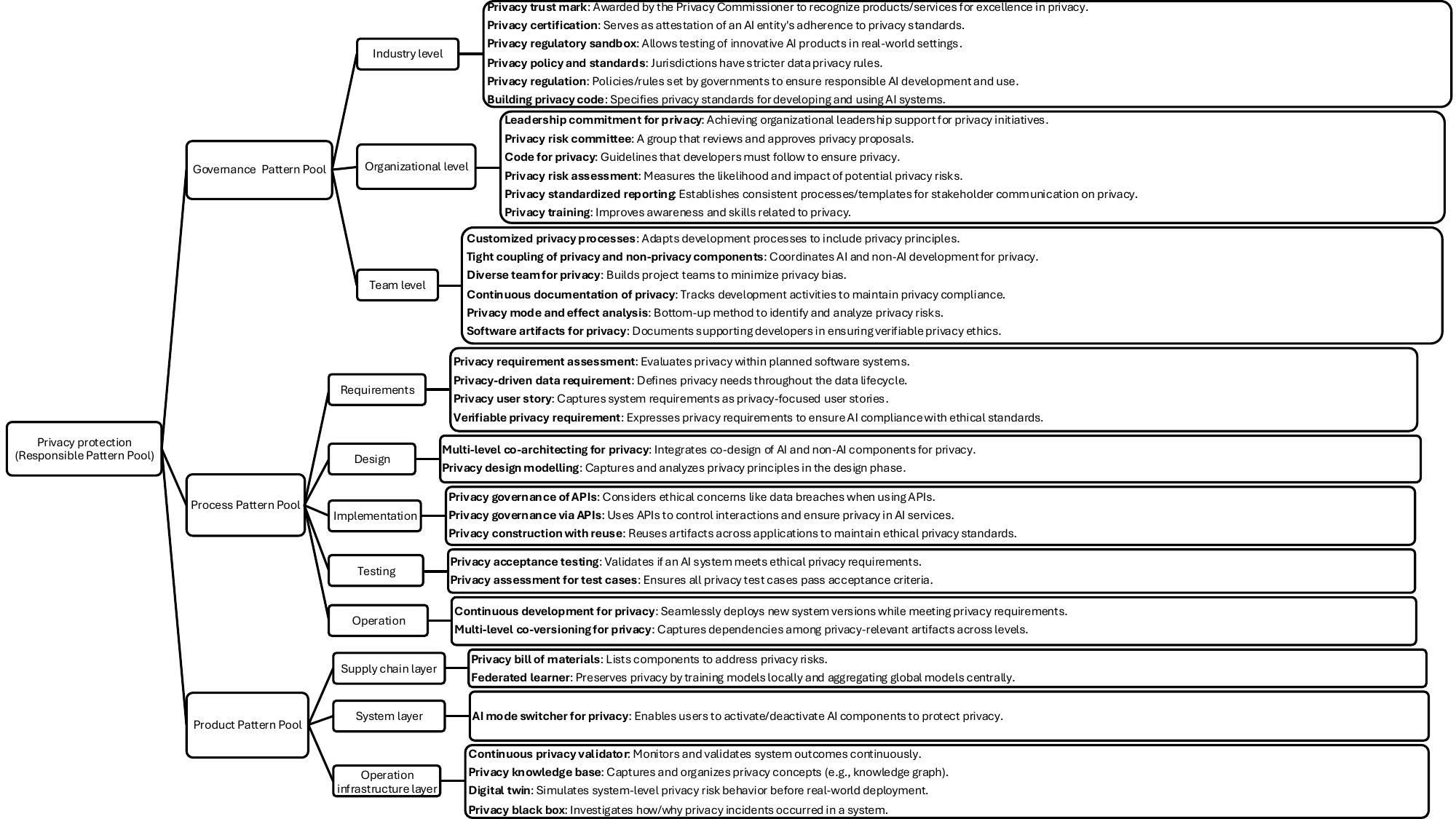}
\vspace{-0.2cm}
\caption{Responsible pattern pool}
\centering
\label{fig_PatternPool}
\end{figure*}

the term ``responsible AI” is used interchangeably with other related terms such as ``ethical AI”, ``trustworthy AI”, ``AI for Good”, ``values-driven AI”, and, more broadly, ``digital humanism”. However, despite their nuances, these terms share a common goal: to promote the development, deployment, and use of AI systems that have a positive impact on individuals, groups, and society while minimizing associated risks.  Based on the existing work~\cite{lu2023towards} , we introduce RAI patterns that are reusable solutions to recurring problems, providing a structured and well-tested foundation for addressing complex issues. In this work, we also categorize these types of patterns into three key domains. We adopt a pattern-oriented approach for evaluating foundation model-based AI systems, particularly privacy. We apply this multi-layer and multi-aspect framework in the following aspects:  
\begin{itemize}
    \item \textbf{Governance pattern}. Governance patterns can be classified into multiple levels (from a \textit{Model Monitoring} perspective in Fig.~\ref{FoundationModel_based_Systems}); for example, organization-level (e.g., privacy risk committee), team level (e.g., how developers perceive privacy), and end-user level (e.g., what end-users concern about privacy).
    
    \item \textbf{Process pattern}. We incorporate process-oriented patterns into the development process (i.e., the development life cycle). We describe each stage of software development life cycle and their potential privacy risks (from a \textit{Model Development} perspective in Fig.~\ref{FoundationModel_based_Systems}). 
    
    \item \textbf{Product pattern}. Product patterns provide reusable solutions and guidance on designing RAI systems, especially privacy. A number of product patterns could be applied to ensure privacy in the foundation model-based systems (from a \textit{Data Management} perspective). 
\end{itemize}

\subsubsection{Evaluation Steps}

specifically, the conceptual framework consists of four steps. In the following, we describe our research agenda in more detail for each step: (1) be aware of privacy requirement and search privacy relevant patterns in the pattern pool (see Fig.~\ref{fig_PatternPool}); (2) interact with stakeholders and suggest patterns to ensure privacy; (3) reflect the used patterns to clarify the design process and stakeholders' demands; and (4) summary, document, and release the outputs for stakeholders. In the following section, we explain the principles of how to apply the framework to ensure privacy, and the detailed pattern pool is shown in Fig.~\ref{fig_PatternPool}.

\smallskip
\noindent\textcolor{teal}{\ding{49}} \textbf{Step-1 (\textit{Searching}). Probing privacy requirements.} 

First, we articulate the privacy goals and their relevant tactics. The structure of privacy protection (see Table \ref{table_privacygoal}) facilitates the determination of the most relevant abstract patterns. These abstract goals can be used as a means around which tactics can be organized in a hierarchical manner and from which context-specific protection goals can be derived. Thus, RAI patterns can be organized in a hierarchical manner to represent both abstract and concrete solutions.

Next, we present the elements of threat scenarios and how they are mapped to the elements of a quality attribute scenario (in Fig. \ref{Our_Framework}). The threat source is the source of a stimulus and the threat event is the stimulus. We subsequently apply the patterns to address the harms. 


\textbf{\textit{Governance patterns}}:
we search governance decisions that involve multi-domain collaboration, including legal, ethical, and data privacy experts, and foster transparency through active stakeholder engagement.
\begin{itemize}
    \item  Break down privacy requirements and assess policies: establish a governance mechanism that involves stakeholders from various levels, including legal, ethics, and data privacy, to identify and classify privacy requirements collaboratively. This pattern ensures that a comprehensive set of privacy concerns is considered, including multiple organizational levels.

    \item Seek stakeholder privacy concern: engage with stakeholders, such as end-users, data subjects, and privacy experts, to gather insights into their privacy expectations and concerns. This pattern fosters transparency in the privacy assessment process. 
    \item Industry level stakeholders and policy: engage with solution procurers; and be aware of privacy regulation, etc;
    \item Organizational level stakeholders and standards: engage with manager and employees; and privacy standardized document patterns.
    \item Team level stakeholders and their background knowledge: engage with privacy expertise developers and end-users; and apply privacy effect analysis patterns.

\end{itemize}

\textbf{\textit{Process patterns}}:
search patterns (i.e., software \textbf{development life cycle}) that ensure traceability of privacy requirements throughout the development life cycle, and systematically search those documentations, and promote interdisciplinary collaboration for actionable solutions.
\begin{itemize}
    \item Traceability and change management: establish a process for tracing privacy requirements throughout the development life cycle and managing changes or updates to these requirements. This pattern ensures that evolving privacy concerns are addressed effectively.
    \item Privacy requirement continually documentation: identify the privacy requirements systematically, including their classification, priority, and relevance to different project phases. This pattern forms the basis for subsequent steps in the evaluation process. Interdisciplinary collaboration pattern: this category of patterns ensures that the identified requirements are actionable and feasible within the project's scope.
\end{itemize}

\textbf{\textit{Product patterns}}:
search and prioritize the reusable solutions in terms of privacy, ensuring alignment between usability and privacy principles right from the design phase.

\begin{itemize}
    \item Product of enhancing user experience: probing product patterns, from \textbf{supply chain, system, and operation infrastructure} layers, that identify privacy requirements. These products ensure that the identified requirements align with the product's usability and functionality, emphasizing a user central approach to privacy.
    \item Privacy products by default: searching and incorporate privacy break-down requirements into the product design usage phase. Ensuring those patterns design usage by default, the system adheres to privacy principles.
\end{itemize}

\find{

\textbf{ Summary of Step-1}

\ding{52}  \ Systematically understand the privacy break-down goals from both data and process orientated perspectives.

\ding{52}  \ Abstract privacy goals articulated what is being protected and who from. It helps stakeholders to understand the context-specific goals and special instances

\ding{52} \ Given the privacy goals, searching governance, process, and product patterns, the first step of framework provides stakeholders with a guidance on how to implement privacy ethical and practices. This helps stakeholders navigate complex ethical terrain and make informed decisions.
}

\smallskip
\noindent\textcolor{teal}{\ding{49}} \textbf{Step-2 (\textit{Suggesting}). Responsible pattern suggestive}. 
We transitioned from classifying privacy requirements in \textbf{Step-1} to proactively suggesting suitable RAI patterns.

\textbf{\textit{Governance patterns}}
review organizational policies for comprehensive privacy protection and implements ethical review mechanisms to assess stakeholder involvement.
\begin{itemize}
    \item Governance pattern review: we review and suggest governance patterns not applied in \textbf{Step-1} focusing on the organizations' policies and \textit{trade-offs }of decision-making for privacy. Suggesting those patterns provides a guideline for establishing robust governance practices reinforcing privacy protection at every organizational level.
    \item Ethical review and approval: we then further suggest a mechanism for conducting ethical review with stakeholders \cite{gurses2016privacy,biempirical,canedo2023privacy}. This ensures that systems involving foundation models undergo thorough privacy and ethical assessments before deployment.
\end{itemize}

\textbf{\textit{Process patterns}}
that advocate for a privacy-by-design approach, ensuring privacy considerations are seamlessly integrated from the project's inception to its completion.
\begin{itemize}
    \item Privacy-by-design workflow: a privacy-by-design workflow is applied to incorporate at this system design stage. This pattern advocates for the continuous evaluation and implementation of privacy measures, ensuring that privacy considerations are integrated from the initial concept to the final product.
\end{itemize}

\textbf{\textit{Product patterns}}
that integrate advanced privacy-enhancing technologies and prioritize user-centric consent mechanisms to align with user autonomy and privacy principles.
\begin{itemize}
    \item Privacy-enhancing technology products: considering and  integrating multi-layered security mechanisms, such as federated learning, differential privacy, software bill of materials (SBOMs), and trust mark, and homomorphic encryption into the design to ensure privacy design.

    \item User-centric consent: evaluating the implementation of user-centric consent mechanisms. Applying pattern products aligning with user autonomy and consent principles.
\end{itemize}

The techniques in Fig. \ref{fig_PatternPool} are considerable as fundamental design decisions (i.e., patterns), and each privacy goal (in Table \ref{table_privacygoal}) may refine different patterns. For example, data minimization, as an abstract tactic, can be refined into concrete RAI patterns. Thus design decisions can be organized in a hierarchical manner to represent both abstract goals and concrete patterns. In order to choose an appropriate RAI pattern, it is important to determine whether the pattern relates to unlinkability, intervanability, or transparency. 

\find{\textbf{Summary of Step-2}

\ding{52}  \ Navigate privacy goals hierarchically and represent the abstract privacy goals and concrete patterns, and the apply RAI patterns to promote privacy by incorporating practices and standards, ensuring that the system meets privacy.
}



\smallskip
\noindent\textcolor{teal}{\ding{49}} \textbf{Step-3 (\textit{Reflecting}). Pattern usage reflective}. Our framework proceeds reflective hints that challenge design decisions and pattern usage in \textbf{Step-1 and Step-2}.  This can be achieved by analyzing privacy RAI patterns with respect to their intent and/or problem, motivation, forces, and context, and solution. 

The underlying abstract privacy goals and the definition of concrete RAI patterns can be mapped to the intent and/or problem of privacy design. By mapping these factors, it is possible to decide whether or not the intent and/or problem of privacy goals abstractly reflects the underlying goals. In addition, the privacy scenarios can be considered as characterizations of privacy requirements as they are generated from a reasonable set of threat scenarios. Based on this, a quality scenario can be used as an indicator to the abstract scenario of a RAI patterns. 

The reflective hints can be inferred from interviews and focus groups. This inference can be made based on reasoning elements. By promoting reflection, this step highlights the potential of our framework in mitigating cognitive biases. 

\textbf{\textit{Governance patterns}} 
that both prompt critical reflection on bias mitigation efforts and emphasize aligning organizational practices with ethical privacy principles.
\begin{itemize}
    \item Bias mitigation review: encourage reflection on the governance mechanisms in place for identifying and mitigating privacy biases. This pattern prompts stakeholders to assess the effectiveness of bias mitigation efforts.
    \item Ethical alignment review: promote reflection on how well the organization's governance practices align with privacy ethical principles and values. This pattern reinforces the necessity of consistently aligning business practices with ethical and legal considerations.
\end{itemize}

\textbf{\textit{Process patterns}} that reviews the importance of iterative, feedback-driven development to ensure privacy measures adapt over time and incorporate valuable stakeholder input. 
\begin{itemize}
    \item Continuous development review: reflect the iterative development and improvement process on privacy. This category of patterns emphasizes the importance of evaluating and adapting privacy measures over time, particularly in response to evolving technologies and user needs.

    \item Feedback-driven development: reflect on the feedback loops and mechanisms in place for gathering user and stakeholder input. In this step, we reflected the value of actively involving end-users and other stakeholders in the privacy design process, ensuring that the techniques and solutions developed are both robust and user-friendly.
\end{itemize}

\textbf{\textit{Product patterns}}
that necessitate reflective evaluation of product pattern impacts, particularly concerning privacy and potential biases, to meet diverse user needs and expectations.
\begin{itemize}
    \item Product pattern impact assessment: reflect on the potential impact of the products on end-users, including privacy concerns and potential biases. This pattern prompts the evaluation of how well the product meets the needs and expectations of diverse user groups.
\end{itemize}

\find{\textbf{Summary of Step-3} 

\ding{52}  \ Review and refine the alignment of abstract privacy goals with the intent and problem of privacy design. This process aids in the assessment of how well the intentions behind RAI patterns correspond to the fundamental privacy goals.
}

\smallskip
\noindent\textcolor{teal}{\ding{49}} \textbf{Step-4. Combination of probing and suggestive outputs}. In assistive privacy elevation, our framework combines probing, suggestive, and reflective steps to tailor the evaluation process based on specific requirements (i.e., privacy). Our framework can potentially provide much more complex knowledge other than patterns in this work, depending on the ontology models' extent. Designing a more comprehensive evaluation framework is however challenging, calling for further research. 

\textbf{\textit{Governance patterns}}
that synthesize initial evaluations into a comprehensive, ethically sound model and employ a continuous feedback loop for dynamic updates.
\begin{itemize}
    \item Holistic governance pattern integration: in the final step, we evaluated the initial results from the various governance-related patterns into a comprehensive governance model. This category of patterns seeks to create a unified governance structure that is ethically sound and aligns with identified privacy requirements.

    \item Governance pattern feedback loop: in this stage, we established a continuous review and update cycle for governance pattern mechanisms. This stage ensures that governance patterns are dynamically updated, based on reflections and insights from the earlier two steps.
\end{itemize}

\textbf{\textit{Process patterns}}
that both integrate various procedural insights into a consolidated, adaptable framework and employ iterative refinement based on reflective feedback and metrics.
\begin{itemize}
    \item Process pattern synthesis: combine procedural insights from privacy requirement documentation, traceability, interdisciplinary collaboration, and privacy-by-design workflows. This pattern aims to construct a consolidated, flexible process that is both robust and adaptable.

    \item Iterative process pattern refinement: utilize reflective feedback and continuous improvement metrics for periodic process updates. This pattern helps in maintaining the effectiveness and relevance of processes over time.
\end{itemize}

\textbf{\textit{Product patterns}}
that fuse elements of user experience, security, and privacy into a unified construct while incorporating periodic re-evaluations based on user impact assessments.
\begin{itemize}
    \item Composite product pattern design: merge elements of user experience, default privacy settings, and security mechanisms into a unified product design. This pattern aspires to strike a balance between user needs, system security, and privacy compliance.

    \item Product pattern re-evaluation: incorporate user impact assessments and feedback-driven insights for periodic product updates. This pattern ensures that the product continually aligns with evolving privacy standards and user expectations.
\end{itemize}

\find{
\textbf{Summary of Step 4}

\ding{52}   \ Expand the existing framework with modular components designed to assess the identified quality attributes. Each component should focus on a particular quality attribute, making it easy to incorporate new attributes into the framework to make decisions.

\ding{52}  \ Enriching the responsible AI pattern pool by incorporating elements from snowboarding and leveraging a knowledge graph can provide valuable insights into the ethical and sustainable development of AI systems..}

\smallskip
\noindent\textcolor{teal}{\ding{171}} To facilitate a guided journey through the evaluation process.
Fig 2 maps privacy goals to RAI patterns through the four evaluation steps, linking each stage to its corresponding patterns (see Fig. \ref{fig_PatternPool}). Columns include ``Evaluation Step'' indicating which framework steps are being considered; ``Responsible Patterns'' categorizing each pattern into governance, process, or product; and ``Output'' summarizing outcomes.

\section{CALL FOR ACTIONS}

\label{sec_analysis}

As we apply the conceptual framework, we tackle challenges that involve consolidating fragmented knowledge, understanding it, and structuring it for assessing privacy in FM-based systems. In this section, we explore the technologies and studies relevant to the framework's attributes.


\smallskip
\noindent\textbf{Generalize the framework to other quality design scenarios}. After applying our framework (Fig. \ref{fig_PatternPool}) to the governance, process, and product patterns, the next frontier is its adaptability to other domains of QAs. This final step serves as a pivot, transitioning the framework from a privacy-centric methodology into a versatile tool that can address various aspects of RAI. Building upon the insights and patterns uncovered and applied in earlier stages, we create templates that can be customized for assessing other quality attributes, such as fairness, accountability, and transparency. This generalization aims to establish a robust and flexible evaluation structure that can be applied across diverse quality domains, enhancing its utility and impact in the broader context of RAI development.

\smallskip
\noindent\textbf{Technologies for enabling the framework}: Collecting scattered design information is challenging. Privacy-relevant information can come from different sources, stakeholders, and multiple standards and guidelines. We tried to apply RAI patterns to cover different aspects, however, during the ``searching" and ``suggesting" phases, consistency pattern checks and case generation were extensive.  In addition, as developers generally do little to organize privacy design, our framework needs to provide automated support to extract the semantics of the information to organize it into knowledge. First, our framework has to understand, in the sense of a machine without comprehension, the meaning of the contents, There are many approaches to interpreting contents, e.g., semantic approaches and NLP semantically analyze the requirements, and requirements can be clustered by different RAI patterns. 

\smallskip
\noindent\textbf{Interactive privacy knowledge inquiry}. Traditional information search engines typically search for information through some word matching or indexing mechanisms. There are two issues with this approach, first, such a search engine does not understand the privacy semantics of a search to help it cut down its search space. Second, such a search engine cannot reason with an inquiry to ask for contextual clarifications. As such, in future work, we advocate an intelligent and interactive inquiring capability. This capability requires the framework to interact with stakeholders to understand what and why a privacy inquiry is about. Stakeholders need to provide more context when promoted, and the framework can then make sure of the additional information to home in its search. In addition, in order to service complex questions, the framework may interpret supportive design contexts, and ask background questions to clarify misunderstandings, ambiguous keywords, or assumptions in terms of privacy.

\section{CONCLUSION}\label{sec_con}
We proopse a conceptual framework for evaluating the privacy sector of foundation model based systems. Our framework offers valuable support to stakeholders seeking to select appropriate patterns to ensure privacy from various perspectives. While the model can be applied across different domains, it primarily targets data-sharing software systems. Pattern selection is guided by contextual factors and their impact on system qualities. The model underwent evaluation and refinement, incorporating input from expert stakeholders. The results validate the model's ability to bring structure and reasoning to the decision-making process.

One of our future goals is to explore additional \textit{trade-off decisions}, including considerations of system usability. We also recommend the development of an \textit{expanded pattern pool} to provide more detailed descriptions of each pattern. In addition, our intention is to create a tool for design guidance, and we invite other researchers to contribute to this pattern tool to facilitate them to design responsible AI systems.

\balance 

\bibliographystyle{IEEEtran}
\bibliography{bib(short)}

\end{document}